# Automatic differentiation for coupled cluster methods

Fabijan Pavošević[*] and Sharon Hammes-Schiffer

Department of Chemistry, Yale University
225 Prospect Street, New Haven, Connecticut 06520 USA

[*]Corresponding author email: fabijan.pavosevic@yale.edu

**Abstract**

Automatic differentiation is a tool for numerically calculating derivatives of a given function up to machine precision. This tool is useful for quantum chemistry methods, which require the calculation of gradients either for the minimization of the energy with respect to wave function parameters or for the calculation of molecular responses to external perturbations. Herein, we apply automatic differentiation to the coupled cluster with doubles method, in which the wave function parameters are obtained by minimizing the energy Lagrangian. The benefit of this approach is that the $\lambda$ amplitudes can be obtained without implementation of the usual $\Lambda$−equations, thereby reducing the coding effort by approximately a factor of two. We also show that the excitation energies at the coupled cluster level can be ontained with only a few lines of the code using automatic differentiation. We further apply automatic differentiation to the multicomponent coupled cluster with doubles method, which treats more than one type of particle, such as electrons and protons, quantum mechanically. This approach will be especially useful for prototyping, debugging, and testing multicomponent quantum chemistry methods because reference and benchmark data are limited in this emerging field.

Many optimization algorithms require the calculation of gradients, which are indispensable for machine learning algorithms such as artificial neural networks, computer vision, and natural language processing.[1] Methods for calculating gradients can be classified into four groups: hand-coded analytical derivatives, symbolic differentiation, numerical differentiation, and automatic differentiation.[2-4] The first method, which requires manual derivation and coding of analytical gradients, is exact and often provides the fastest code, but it can be labor intensive and error prone. While symbolic differentiation is exact and provides a symbolic expression, it can lead to a memory intensive and inefficient code. Numerical differentiation is easy to code and is very useful as a debugging tool, but it is prone to instabilities due to round-off errors and is slow in high dimensions. The fourth approach, automatic differentiation (AD),[5] is exact up to machine precision with speed comparable to that of hand-coded analytical derivatives. In the AD approach, numerical values of derivatives to an arbitrary order are calculated by repeatedly applying the chain rule to elementary arithmetic operations and elementary functions. Unlike symbolic differentiation, AD does not produce a symbolic expression for the derivative but rather produces a numerical value of the derivative with the computational cost determined by the evaluation of the input function. AD is a state-of-the-art tool in machine learning algorithms[6] that is available in the TensorFlow software library[7] developed by Google Brain and in the PyTorch library[8] developed by Facebook's artificial intelligence research lab, in addition to other successful implementations.[9-12]

Many electronic structure methods in quantum chemistry rely on minimization of the energy with respect to wave function parameters, such as the Hartree-Fock method or the coupled cluster approach.[13-15] Furthermore, calculations of molecular response properties, such as magnetizability, dipole moment polarizability, or NMR parameters, require the calculation of derivatives with respect to external fields.[16, 17] Typically, these quantities are calculated analytically, and quantum chemists invest a considerable amount of time and effort in deriving and implementing these equations. The potential of AD for resolving these problems has been recognized by the quantum chemistry community. The pioneering work by Thiel and coworkers employed an AD algorithm to generate analytical gradient code for several semiempirical self-consistent field methods.[18] Subsequently, the AD algorithm was used for the calculation of derivatives up to an arbitrary order of exchange-correlation energy functionals in density functional theory.[19] More recently, AD was implemented in the *DiffiQult* software package, and its capability was demonstrated on the fully variational Hartree-Fock method.[20] Neuscamman's group has also employed AD in the context of quantum Monte Carlo[21] and for minimization of the excited state mean-field theory Lagrangian using TensorFlow.[22, 23]

Encouraged by the promise of AD, in this work we investigate the potential benefit of AD for the evaluation of the CC amplitudes that parametrize the wave function. As a proof of concept, we use AD to implement the coupled cluster with doubles (CCD) method. The main goal of this paper is to show that the CC amplitudes can be obtained with reduced amount of coding effort using AD. Another goal of this paper is to highlight the benefits of AD in the context of multicomponent CC methods for treating more than one type of particle quantum mechanically, such as electrons and protons. As a proof of concept, we use AD to implement the CCD method within the nuclear-electronic orbital (NEO) framework. AD is particularly useful in this context because the NEO-CCD method requires a much greater number of terms than its conventional CCD counterpart.

The CC approach offers a robust and accurate strategy for solving the Schrödinger equation.[14, 15, 24] In this approach, the wave function is represented by the exponentiated excitation cluster operator that includes correlation effects between electrons through single, double, triple, and higher excitations from the reference determinant. The coupled cluster energy is calculated from the energy Lagrangian as

$$E_{CC} = \langle 0|(1+\hat{\Lambda})e^{-\hat{T}}\hat{H}e^{\hat{T}}|0\rangle \quad (1)$$

This equation is a starting point for the calculation of various molecular properties at the CC level.[17] In this equation, $\hat{H} = h_q^p a_p^q + \frac{1}{4}\bar{g}_{rs}^{pq} a_{pq}^{rs}$ is the electronic Hamiltonian, where $h_q^p \equiv \langle q|\hat{h}^e|p\rangle$ and $\bar{g}_{rs}^{pq} \equiv \langle rs\|pq\rangle$ are matrix elements of the core electronic Hamiltonian and an antisymmetrized two-electron repulsion tensor element, respectively. Additionally, $a_{p_1 p_2 \ldots p_n}^{q_1 q_2 \ldots q_n} = a_{q_1}^\dagger a_{q_2}^\dagger \ldots a_{q_n}^\dagger a_{p_n} \ldots a_{p_2} a_{p_1}$ are second-quantized electronic excitation operators expressed in terms of creation/anihilation ($a_p^\dagger / a_p$) fermionic operators. The occupied electronic spin orbitals are denoted by $i,j,k,l,\ldots$, the unoccupied spin orbitals are denoted by $a,b,c,d,\ldots$, and the general electronic spin orbitals are denoted by $p,q,r,s,\ldots$. The Einstein summation convention over repeated indices is assumed thoughout this Communication. The ket $|0\rangle$ is the reference wave function, which is usually the Hartree-Fock (HF) wave function.

In Eq. (1), the operators $\hat{T} = t_\mu a^\mu$ and $\hat{\Lambda} = \lambda^\mu a_\mu$ are the excitation and de-excitation cluster operators, where $a^\mu = a_\mu^\dagger = \{a_i^a, a_{ij}^{ab}, a_{ijk}^{abc}, \ldots\}$ is a set of single, double, triple and higher excitation operators and $\mu$ is an excitation manifold. The parameters $t_\mu$ and $\lambda^\mu$ are unknown amplitudes that are determined by optimization of Eq. (1) with respect to $\lambda^\mu$ and $t_\mu$, respectively:

$$\frac{\partial E_{CC}}{\partial \lambda^\mu} = \langle 0|a_\mu e^{-\hat{T}}\hat{H}e^{\hat{T}}|0\rangle = {}^t\sigma_\mu = 0 \quad (2)$$

$$\frac{\partial E_{CC}}{\partial t_\mu} = \langle 0|(1+\hat{\Lambda})[e^{-\hat{T}}\hat{H}e^{\hat{T}}, a^\mu]|0\rangle = {}^\lambda\sigma^\mu = 0 \qquad (3)$$

Eqs. (2) and (3) are known as the $t$–amplitude equations and the $\Lambda$–equations, respectively. Finally, the excitation energies at the coupled cluster level are obtained by diagonalizing the Jacobian matrix defined by

$$\frac{\partial {}^t\sigma_\mu}{\partial t_\nu} = \langle 0|a_\mu[e^{-\hat{T}}\hat{H}e^{\hat{T}}, a^\nu]|0\rangle = J_\mu^\nu \qquad (4)$$

Truncation of the excitation cluster operator $\hat{T}$ up to a certain excitation rank establishes the hierarchy of the CC methods. The simplest CC method that incorporates correlation effects between electrons is the coupled cluster with doubles (CCD) method, in which the cluster excitation and de-excitation operators are $\hat{T}_2 = t_{ab}^{ij} a_{ij}^{ab}$ and $\hat{\Lambda}_2 = \lambda_{ij}^{ab} a_{ab}^{ij}$. Even for the CCD method, the derivation and implementation of Eqs. (2) and (3) into a quantum chemistry program require a considerable amount of effort that grows rapidly for the CC methods with higher excitation ranks.[15, 24]

An alternative route for the calculation of the same $t_{ab}^{ij}$ and $\lambda_{ij}^{ab}$ amplitudes is as follows: first construct the Lagrangian given in Eq. (1) and then perform AD on this Lagrangian according to Eqs. (2) and (3). Because the Lagrangian in Eq. (1) is constructed by augmenting the CC energy $\langle 0|e^{-\hat{T}_2}\hat{H}e^{\hat{T}_2}|0\rangle = E_{HF} + \frac{1}{4}\bar{g}_{ij}^{ab} t_{ab}^{ij}$ with the $t$–amplitude equations (Eq. (2)) weighted by the Lagrange multipliers $\lambda_{ij}^{ab}$, implementation of the Lagrangian requires implementation of the $t$–amplitude equations in addition to very minor modifications (see Figure 1). Once the Lagrangian is available, the $t_{ab}^{ij}$ and $\lambda_{ij}^{ab}$ amplitudes are calculated with AD. Therefore, an advantage of calculating the amplitudes via AD is that it does not require implementation of the Λ−equations, thus reducing the coding effort by approximately a factor of two.

```
1   with tensorflow.GradientTape(persistent=True) as t:
2       t.watch((l2_ijab, t2_abij))
3       t2_sigma = get_t2_sigma(f_mo, g_mo, t2_abij)
4       L  =     tf.einsum('ijab,abij ->', l2_ijab, t2_sigma)
5       L += 1/4*tf.einsum('ijab,abij ->', g_ijab, t2_abij)
6
7   dL_dl, dL_dt = t.gradient(L, (l2_ijab, t2_abij))
8
9   t2_abij += tf.einsum('iajb,ijab -> abij', e_iajb, dL_dl)
10  l2_ijab += tf.einsum('iajb,abij -> ijab', e_iajb, dL_dt)
```

Figure 1. A piece of code for the calculation of the $t_{ab}^{ij}$ and $\lambda_{ij}^{ab}$ amplitude updates using AD implemented in TensorFlow v2.1.0.

To illustrate these advantages, we have implemented the CCD Lagrangian, $t$−amplitude equations, and Λ−equations in the Psi4Numpy[25] quantum chemistry software. The programmable $t$−amplitude equations and Λ−equations were derived with the SeQuant package. Automatic differentiation was performed using the TensorFlow v2.1.0 program.[7] Tensor contractions and factorizations of expressions that occur in the CCD Lagrangian, $t$−amplitude equations, and Λ−equations were performed with the einsum function implemented in the TensorFlow v2.1.0. The CCD calculations reported in this work were conducted on a benzene molecule using the cc-pVDZ basis set,[26] consisting of 114 orbitals and 42 electrons. The calculations were performed on a MacBook Pro with a 3.1 GHz Intel Core i5 processor and with 8 Gb of memory using a single core.

A piece of the code for the calculation of the $t_{ab}^{ij}$ and $\lambda_{ij}^{ab}$ amplitude updates using AD implemented in TensorFlow v2.1.0 is shown in Figure 1. This piece of code indicates that the Lagrangian (Eq. (1)), denoted L in Figure 1, is constructed in lines 3, 4, and 5: $^t\sigma_{ab}^{ij}$ is constructed from the $t$−amplitude equations in line 3, $^t\sigma_{ab}^{ij}$ is multiplied by $\lambda_{ij}^{ab}$ in line 4, and finally the CCD energy is added in line 5 as $\frac{1}{4}\bar{g}_{ij}^{ab}t_{ab}^{ij}$. Therefore, construction of the Lagrangian requires minor coding effort when the $t$–amplitude equations are available. Once the Lagrangian is constructed, the derivatives defined by Eqs. (2) and (3) are obtained by a single line of code (line 7). The last two lines (9 and 10) calculate the updated $t_{ab}^{ij}$ and $\lambda_{ij}^{ab}$ amplitudes by dividing the derivatives defined in Eqs. (2) and (3) by $\varepsilon_i - \varepsilon_a + \varepsilon_j - \varepsilon_b$, where $\varepsilon$ are orbital energies, according to standard procedure in CC theory.[14, 15, 24] Note that the step defined by Eq. (2) (line 7) is redundant because this information is calculated in line 3, but we include it here for completeness.

Both of these implementations, namely calculating the $t_{ab}^{ij}$ and $\lambda_{ij}^{ab}$ amplitudes by implementing the $t$–amplitude equations and $\Lambda$–equations (denoted conventional herein) or by performing AD on the Lagrangian, should produce the same result. For validation, we calculated both the CCD correlation energy, $E_{\text{CCD}} = \frac{1}{4}\bar{g}_{ij}^{ab}t_{ab}^{ij}$, and the pseudo correlation energy, $E_{\text{pCCD}} = \frac{1}{4}\bar{g}_{ab}^{ij}\lambda_{ij}^{ab}$. Note that the pseudo correlation energy is defined to be the correlation energy calculated with the $\lambda_{ij}^{ab}$ amplitudes. The correlation and pseudo correlation energies calculated with both approaches agree to machine precision, although only the first ten decimal places are shown in Table I. This agreement provides confirmation that both approaches are implemented correctly.

Table I. Timings[a] (in seconds per iteration) for the computation of different contributions and the correlation and pseudo correlation energies (in units of a.u.) in the CCD method with conventional and AD approaches.

|  | Conventional | Automatic differentiation |
|---|---|---|
| $t_{ab}^{ij}$/sec | 12.4 | 26.3 |
| $\lambda_{ij}^{ab}$ /sec | 18.2 | 34.6 |
| Lagrangian /sec | 10.6 | 10.6 |
| $E_{\text{CCD}}$/a.u. | -0.8332168437 | -0.8332168437 |
| $E_{\text{pCCD}}$/a.u. | -0.8054038137 | -0.8054038137 |

[a]Reported timings are for the spin-free implementations.

The average times for calculation of the $t_{ab}^{ij}$ and $\lambda_{ij}^{ab}$ amplitudes with the conventional approach are 12.4 and 18.2 seconds per iteration, respectively (Table I). The computation time for calculation of the $\lambda_{ij}^{ab}$ amplitudes is longer due to the greater number of terms that appear in the Λ−equations. The average times for calculation of these quantities using AD are 26.3 and 34.6 seconds per iteration, respectively (Table I), including the time of 10.6 seconds for the calculation of the Lagrangian. Therefore, once the Lagrangian is calculated, the times for calculation of the derivatives with respect to $\lambda_{ij}^{ab}$ and $t_{ab}^{ij}$ to obtain $t_{ab}^{ij}$ and $\lambda_{ij}^{ab}$, respectively, are 15.7 and 24.0 seconds. The AD computation time for calculating $\lambda_{ij}^{ab}$ is longer than that for calculating $t_{ab}^{ij}$ because the Lagrangian has quadratic dependence on $t_{ab}^{ij}$ and linear dependence on $\lambda_{ij}^{ab}$. The timings given in Table I indicate that the calculation of $t_{ab}^{ij}$ and $\lambda_{ij}^{ab}$ is roughly two times longer with AD than with the conventional approach for this system. However, the advantage of the AD approach is that it does not require implementation of the Λ−equations, thereby reducing the coding effort by a factor of two.

In addition to the ground state properties, the excitation energies at the CCD level of theory are the eigenvalues of the Jacobian matrix defined in Eq. (4). Therefore, the AD approach allows an elegant implementation for the excited states methods using the coupled cluster ansatz in only a few lines of the code. The drawback of this approach is that it requires construction of the full

Jacobian matrix thus it can be used only for small systems. Nevertheless, this approach provides a simple method for debugging existing codes and for prototyping a novel methods.

The AD approach can be especially beneficial for prototyping, debugging, and testing of multicomponent CC methods,[27-33] in which more than one type of particle is treated quantum mechanically.[34] In the nuclear electronic orbital (NEO) approach, specified nuclei in addition to all electrons are treated quantum mechanically using molecular orbital techniques. The CCD approach within the NEO framework, denoted NEO-CCD, is a multicomponent generalization of CCD. Within the Lagrangian formalism, the NEO-CCD energy Lagrangian is given by

$$E_{\text{NEO-CCD}} = \left\langle 0^e 0^p \left| \left(1+\hat{\Lambda}_2\right) e^{-\hat{T}_2} \hat{H}_{\text{NEO}} e^{\hat{T}_2} \right| 0^e 0^p \right\rangle \tag{5}$$

where $\hat{H}_{\text{NEO}} = h_q^p a_p^q + \frac{1}{4}\bar{g}_{rs}^{pq} a_{pq}^{rs} + h_Q^P a_P^Q + \frac{1}{4}\bar{g}_{RS}^{PQ} a_{PQ}^{RS} - g_{qQ}^{pP} a_{pP}^{qQ}$ is the NEO Hamiltonian and $\left|0^e 0^p\right\rangle$ is the reference NEO-HF wave function. The upper case indices denote protonic spin orbitals and are defined analogously to the previously introduced electronic lower case indices. Analogous to the electronic counterparts, $h_Q^P \equiv \left\langle Q|\hat{h}^p|P\right\rangle$ and $\bar{g}_{RS}^{PQ} \equiv \left\langle RS \| PQ\right\rangle$ are matrix elements of the core protonic Hamiltonian and an antisymmetrized two-proton repulsion tensor element, respectively, and $g_{qQ}^{pP} = \left\langle qQ|pP\right\rangle$ is an electron-proton attraction tensor element. Finally, $\hat{T}_2 = \frac{1}{4}t_{ab}^{ij} a_{ij}^{ab} + \frac{1}{4}t_{AB}^{IJ} a_{IJ}^{AB} + t_{aA}^{iI} a_{iI}^{aA}$ and $\hat{\Lambda}_2 = \frac{1}{4}\lambda_{ij}^{ab} a_{ab}^{ij} + \frac{1}{4}\lambda_{IJ}^{AB} a_{AB}^{IJ} + \lambda_{iI}^{aA} a_{aA}^{iI}$ are excitation and de-excitation cluster operators, respectively. Note that $\{a_{ij}^{ab}, a_{IJ}^{AB}, a_{iI}^{aA}\}$ are the second-quantized electron-electron, proton-proton, and electron-proton double excitation operators. The $\{t_{ab}^{ij}, t_{AB}^{IJ}, t_{aA}^{iI}\}$ and $\{\lambda_{ij}^{ab}, \lambda_{IJ}^{AB}, \lambda_{iI}^{aA}\}$ are unknown wave function parameters that are determined from Eqs. (2) and (3), respectively.

The coding effort for implementing the NEO-CCD method compared to the conventional electronic CCD method is much greater because of the larger number of $t$–amplitude equations and $\Lambda$–equations. As an example, the number of terms in the $t$–amplitude equations is 18 for the conventional electronic CCD method and 86 for the NEO-CCD method, and the number of terms in the $\Lambda$–equations is 25 for the conventional electronic CCD method and 123 for the NEO-CCD method. There are approximately five times more terms in the NEO-CCD method that need to be implemented, and this ratio grows rapidly for the higher-rank NEO-CC methods.

We applied the AD approach in the context of the NEO-CCD method. The calculations were conducted on the HCN molecule, with a CN distance of 1.148 Å and a CH distance of 1.058 Å, using the cc-pVDZ electronic basis set[26] and the PB4-F2a′ nuclear basis set.[35] Both the conventional and the AD approaches produce the same values of −0.3032785471 and −0.2939959564 for the correlation energy and the pseudo correlation energy, respectively. This agreement provides confirmation that both approaches are implemented correctly. Because of the far greater number of terms that appear in the $t$–amplitude equations and $\Lambda$–equations within the NEO-CCD method, the AD approach is very attractive. As an example, we previously developed the coupled cluster with singles and doubles method within the NEO framework, denoted NEO-CCSD.[30, 31, 34] Initially, we derived and implemented the $t$–amplitude equations,[30] and after a significant amount of additional effort, we derived and implemented the $\Lambda$–equations for the NEO-CCSD method.[31] The additional effort was substantial because the NEO-CCSD equations include approximately five times more terms than their conventional electronic CCSD counterparts.[30, 31] As shown here, AD allowed us to quickly calculate the $\lambda$ amplitudes with only minor coding effort after the $t$–amplitude equations were implemented.

This work presents the use of AD for the calculation of the $t_{ab}^{ij}$ and $\lambda_{ij}^{ab}$ amplitudes within the CC framework. Although the computation time is roughly two times longer than that for the conventional implementation with analytical gradients for the system studied, this approach has the significant advantage that it does not require implementation of the Λ−equations. The derivation and implementation of the Λ−equations is often challenging because of the greater number of terms relative to the *t*−amplitude equations, and therefore AD reduces the coding effort by approximately a factor of two. Furthermore, AD allows for a quick implementation of the excited states procedures at coupled cluster level of theory. The AD approach is particularly appealing for the development of multicomponent methods, in which more than one type of particle is treated quantum mechanically, because of the dramatic increase in the number of terms in multicomponent CC methods compared to their conventional electronic counterparts.[33, 36-40] In general, the AD approach can be used for prototyping, debugging, and testing methods for which high level reference results are limited, with the emerging field of multicomponent quantum chemistry serving as a compelling example. Furthermore, AD may be useful for the calculation of various molecular response properties at the CC level. Thus, this work provides the foundation for a wide range of future methodological developments and applications of AD in quantum chemistry.

**Acknowledgement**

The authors thank Dr. Luning Zhao, Dr. Tanner Culpitt, Dr. Qi Yu, Dr. Saswata Roy, Zhen Tao, Benjamin Rousseau, Patrick Schneider, and Prof. John Tully for useful discussions. This work was supported by the U.S. Department of Energy, Office of Science, Offices of Basic Energy

Sciences and Advanced Scientific Computing Research, Scientific Discovery through Advanced Computing (SciDAC) program.

**Data Availability Statement**

The data that support the findings of this study are available within the article.